\documentclass[aps,twocolumn,
 superscriptaddress,amsmath,amssymb,nofootinbib]{revtex4-2}
\usepackage{amsfonts}
\usepackage{amsmath}
\usepackage{bbm}
\usepackage{amssymb}
\usepackage{graphicx}%
\usepackage{footmisc}
\usepackage{bm}
\usepackage{braket}
\usepackage{CJK}
\usepackage{hyperref}
\usepackage{cleveref}
\usepackage{longtable,booktabs}

\newcommand{\abs}[1]{\lvert#1\rvert}
\newcommand{\norm}[1]{\lVert#1\rVert}
\newcommand{\Tr}{\text{Tr}}

\begin{document}
\begin{CJK*}{UTF8}{gbsn}
\title{Subsystem eigenstate thermalization hypothesis for translation invariant systems}

\author{Zhiqiang Huang (黄志强)}
\email{hzq@wipm.ac.cn}
\affiliation{State Key Laboratory of Magnetic Resonance and Atomic and Molecular Physics, Innovation Academy for Precision Measurement Science and Technology, Chinese Academy of Sciences, Wuhan 430071, China}
\author{Xiao-Kan Guo (郭肖侃)}
\email{kankuohsiao@whu.edu.cn}
\affiliation{Department of Applied Mathematics, Yancheng Institute of Technology, Jiangsu 224051, China}
\date{\today}

\begin{abstract}
The  eigenstate thermalization hypothesis  for translation invariant quantum spin systems  has been proved recently by using random matrices. In this paper, we study the subsystem version of eigenstate thermalization hypothesis for translation invariant quantum systems without referring to random matrices. We first find a relation between the quantum variance and the Belavkin-Staszewski relative entropy. Then, by showing the small upper bounds on the quantum variance  and the Belavkin-Staszewski relative entropy, we prove the subsystem  eigenstate thermalization hypothesis for translation invariant quantum systems with an algebraic  speed of convergence in an elementary way. The proof holds for most of the translation invariant quantum lattice models with exponential or algebraic decays of correlations.
\end{abstract}


\maketitle

\section{Introduction}\label{INTRO}
The equilibration and the thermalization of an isolated quantum system are fundamental for understanding the emergence of quantum statistical mechanics  from unitary quantum mechanics. By thermalization, it means that either an isolated quantum system  would evolve into a thermal state, or the observables would attain their values in a statistical ensemble, after a unitary quantum evolution of the isolated quantum system for  a period of time that is long enough. Since a unitary quantum evolution preserves the pure state, it is not easy to understand how the statistical mixture emerges  if the initial state of an isolated quantum system is a pure state. Numerous approaches have been proposed to understand various aspects of this problem, cf. the reviews \cite{GE16,MIKU18}.

The eigenstate thermalization hypothesis (ETH) \cite{Rigol16, Deu18}, that the expectation values of quantum observables in an energy eigenstate should approximately coincide with the thermal expectation values,
 provides a possible mechanism for the thermalization of an isolated quantum system. Although the ETH   has more and more numerical and experimental evidences in specific closed quantum models/systems, its physical origin and mathematical description  are not completely understood by now. In the original proposal by Deutsch and  Srednicki \cite{Deu91,Sre94,Sre99}, a random perturbation is added to a closed quantum system, and the ETH holds if the perturbed system becomes chaotic. By modeling the random perturbations  as random matrices, the ETH for deterministic observables with the Hamiltonians sampled from the Wigner random matrix ensemble without further unitary symmetry is mathematically proved  in the recent work \cite{E1}. This scenario, however, is not universal. For one thing, if further unitary symmetries are present, the conserved quantities would obstruct the thermalization to Gibbs states and the original ETH would fail. More recently in \cite{SE}, the ETH for translation invariant spin systems is proved using the same method from random matrices, thereby generalizing its validity  to various translation invariant lattice spin models.

In many studies of the ``weak'' ETH, for example, \cite{BKL10,Mori,IKS17,KS20,MB21}, one does not presume the random energy perturbations, or simply the random Hamiltonians, but tries to derive the statistical properties solely from quantum properties. From this perspective, the quantum entanglement inside a closed quantum system, together with its dynamics under the global unitary evolution, should play a crucial role  for thermalization, which has indeed been experimentally observed in \cite{Kaufman16}. 
To quantify the entanglement in a closed quantum system, we need to work at the level of subsystems of the total  system to compute the entanglement entropies and alike. This observation leads to the {\it subsystem} ETH \cite{DLL18,LDL18}, which hypothesizes the convergence of  the subsystem density matrices to the thermal Gibbs density matrix. In fact, the trace distance between two density matrices is bounded by the relative entropy between two density matrices. Since the entanglement entropies and relative entropies are calculable in many conformal field theories (CFT), the subsystem ETH and its violation have been tested in  many CFTs \cite{LDL18,LDL18b,BDDP17,HLZ17a,HLZ17b,FW18,GLZ19}. Notice that the conformal symmetry forms an infinite-dimensional group, so the infinite number of conserved KdV charges make the generalized Gibbs states as the proper equilibrated states for CFTs \cite{DP19a,DP19b}. It is then natural to ask for a quantum system/model with a smaller symmetry group such that the subsystem ETH  still holds. 

For translation invariant quantum lattice  systems, we already know that the strong ETH \cite{SE}, the weak ETH \cite{Mori,IKS17}, and the canonical typicality \cite{MAMW15} are true. In addition,  a version of the generalized ETH, i.e. thermalization to the generalized Gibbs ensemble, for translation invariant quasi-free fermionic  integrable models is also  proved in \cite{RM18}. We therefore see that the translation invariant quantum lattice systems are good tested for checking various versions of  ETH. In  this paper, we make an effort to  prove the subsystem ETH for translation invariant systems without referring to random matrices.

We will work in the setting of translation invariant quantum lattice system in the sense of \cite{IKS17}.  Unlike the considerations by Iyoda {\it et al.} \cite{IKS17}, we find a formal relation between the quantum variance and the Belavkin-Staszewski relative entropy in an average sense, thereby establishing a connection of the scaling analysis on the variance given in \cite{IKS17} and the subsystem ETH formulated as the relative-entropic bounds on the trace distance between the subsystem state and the canonical thermal state. In fact, we are able to prove the following form of subsystem ETH,
\begin{align}
&\norm{\rho_\text{sub}-\rho^\text{c}_{A}}\sim\mathcal{O}(N^{1/2}_A/N^{1/2}),\label{111}\\
&\norm{\sigma_\text{sub}}\sim \mathcal{O}(N^{1/2}_A/N^{1/2}),\label{222}
\end{align}
where $\rho_\text{sub}$ is the state of a subsystem $A$, $\sigma_\text{sub}$ is a traceless (or ``off-diagonal'')  matrix  of a subsystem, $\rho^\text{c}_A$ is the reduced density matrix of canonical thermal state,
for translation invariant quantum lattice systems.
Notice that in our results \eqref{111} and \eqref{222} the errors decay algebraically as $\mathcal{O}(N^{1/2}_A/N^{1/2})$ with $N_A$ the degrees of freedom (or number of lattice sites) in the subsystem $A$ and $N$ the total degrees of freedom.
This decaying behavior is weaker than the exponential decays as usually expected in ETH but corroborates the algebraic decay of error terms  in the random-matrix proof of ETH for translation invariant systems \cite{SE}.

We begin in Section \ref{II} with some preliminary results about ETH, subsystem ETH, and in particular the setting of translation invariant quantum lattice system from \cite{IKS17}. In Section \ref{III}, we introduce the main technical input, i.e. the formal relation between the quantum variance and the Belavkin-Staszewski relative entropy in an average sense. Using this relation, we analyze the scaling of both the variance and the Belavkin-Staszewski relative entropy  and prove the subsystem ETH in Section \ref{IV}.  In Section \ref{5V}, we discuss the role of correlation decay in our proof. In the final Section \ref{CD} we conclude this paper and discuss some related issues.
\section{Preliminaries}\label{II}
In this section, we recollect the basics of ETH and subsystem ETH, and the weak ETH with eigenstate typicality in the sense of \cite{IKS17}.
\subsection{ETH and subsystem ETH}
Consider an isolated or closed quantum system $B$ with Hamiltonian $h$. This Hamiltonian $h$ could include a random perturbation $h_\text{pert.}$. Suppose $h$ has  eigenvectors $\ket{E_i}, i=1,2,...,N$ with energy eigenvalues $E_i$, i.e. $h\ket{E_i}=E_i\ket{E_i}$. For a few-body observable $A$, the local  ETH can be formulated in terms of the expectation values of $A$ in the energy eigenstates as 
\begin{equation}\label{1}
\braket{E_i|A|E_j}=\mathcal{A}(E)\delta_{ij}+e^{-S(E)/2}f(E,\omega)R_{ij}
\end{equation}
where $E=\frac{1}{2}(E_i+E_j)$, $\omega=E_i-E_j$, and $e^{S(E)}=E\sum_i\delta(E-E_i)$ is the  density of states of the system $B$. The $\mathcal{A}(E)$ and $f(E,\omega)$ are smooth functions, while the fluctuation  factor $R_{ij}$ is of order $1$. Particularly the thermalization requires that  $\mathcal{A}(E)$ should be approximately the thermal average of $A$ in the canonical ensemble, $\mathcal{A}=\braket{A}_\text{c}+\mathcal{O}(N^{-1})+\mathcal{O}(e^{-S/2})$, in the large $N$ limit.

This local form \eqref{1} of ETH can be derived based on Berry's chaotic conjecture \cite{Sre99}. If we sample the Hamiltonian $h$ from a random matrix ensemble, the  following form of inequality for ETH,
\begin{equation}\label{2}
\abs{\braket{E_i|A|E_j}-\braket{A}_\text{mc}(E)\delta_{ij}}\leqslant \mathcal{O}(e^{-S/2}),
\end{equation}
where $\braket{}_\text{mc}$ denotes the thermal average in the microcanonical ensemble,
 can be proved mathematically in several cases, including the translation invariant systems, by using properties of random matrices \cite{E1,SE}. 

Both \eqref{1} and \eqref{2} are  local conditions, as the ETH are assumed for each energy eigenstate. Therefore, in analogy to the canonical typicality of a subsystem $B_1$,\footnote{We emphasize that, throughout this paper, $B$ without indices denotes the total system and $B_i$ and likewise denote the subsystems.} we can envision the {\it subsystem} ETH,
\begin{align}
\norm{\rho^{B_1}_i-\rho^\text{c}(E_i)}&\sim\mathcal{O}(e^{-S/2}),\label{3}\\
\norm{\rho^{B_1}_{ij}}&\sim\mathcal{O}(e^{-S/2}), \quad i\neq j\label{4}
\end{align}
where $\rho^{B_1}_i=\text{Tr}_{\bar{B}_1}\ket{E_i}\bra{E_i}$ is the reduced density matrix of the subsystem $B_1$,
$\rho^\text{c}$ is a universal density matrix that could be the thermal canonical one, and $\rho^{B_1}_{ij}=\text{Tr}_{\bar{B}_1}\ket{E_i}\bra{E_j}$. The norm here refers to the trace distance, or Schatten $1$-norm, $\norm{\rho_1-\rho_2}=\frac{1}{2}\text{Tr}\sqrt{(\rho_1-\rho_2)^2}$. The subsystem ETH as  given by \eqref{3} and \eqref{4} is in fact stronger than the local ETH as in \eqref{1}, due to the following inequality \cite{LDL18}
\begin{equation}
\abs{\braket{A}-\braket{A}_\text{c}}\leqslant\sqrt{\norm{\rho-\rho^\text{c}}\text{Tr}[(\rho+\rho^\text{c})A^2]}
\end{equation}
where $\braket{A}=\text{Tr}(\rho A)$ and $\braket{A}_\text{c}=\text{Tr}(\rho^\text{c} A)$. 

What is important in the following is that the trace distance in \eqref{3} can be  bounded by the relative entropy between  two density matrices,
\begin{equation}\label{6}
\norm{\rho^{B_1}-\rho^\text{c}(E_i)}^2\leqslant 2S(\rho^{B_1}||\rho^\text{c}),
\end{equation}
where $S(\rho_1||\rho_2)=\text{tr}(\rho_1\log\rho_1)-\text{tr}(\rho_1\log\rho_2)$ is the (Umegaki) quantum relative entropy. This inequality \eqref{6} is the so-called quantum Pinsker inequality  in quantum information theory \cite{Wat18}.

\subsection{Weak ETH with eigenstate typicality}\label{IIB}
In proving the  weak ETH for translation invariant quantum lattice systems \cite{IKS17}, the quantum uncertainty of measuring an observable plays an important role. Conventionally, the uncertainties, either classical or quantum, can be quantified by the variance \cite{L05}. For instance, given a quantum state $\rho$, the quantum uncertainty of measuring an observable $A$ in the state $\rho$ can be quantified by the variance
\begin{equation}\label{VAR}
    V(\rho,A)=\Tr (\rho A A^\dag)-\abs{\Tr\rho A}^2=\Tr [\rho(A-\braket{A})(A-\braket{A})^\dag].
\end{equation}
Let $\rho_B=\sum_j p_j \Pi^j_{B}$ be the state of the total system $B$ expanded in the orthonormal basis $\{\Pi^j_{B}\}$ of rank-$1$ projectors, then in terms of these projectors one can define particularly the following quantity, which is called as  {\it fluctuation} in \cite{IKS17},
\begin{align}\label{FLUC}
    \Delta(\rho,A)=\sum_j p_j \abs{\Tr \Pi^j_{B} A}^2-\abs{\Tr\rho A}^2.
\end{align}
We have
\begin{align}\label{FLUCBVAR}
    \Delta(\rho,A)\leqslant \sum_{j,k} p_j \Tr (\Pi^j_{B} A \Pi^k_{B}A^\dag) -\abs{\Tr\rho A}^2=  V(\rho,A),
\end{align}
because   the additional off-diagonal terms are positive, i.e.
$\Tr (\Pi^j_{B} A \Pi^k_{B}A^\dag)=\abs{\bra{j}A\ket{k}}^2\geqslant 0$.
This $\Delta(\rho,A)$ is related to the following (in)distinguishability  measure of quantum states:
\begin{equation}\label{IDUMO}
    d(\Pi^j_{B},\rho;A)=\abs{\Tr [(\Pi^j_{B} -\rho) A]}^2.
\end{equation}
Indeed, $\Delta(\rho,A)$ can be considered as the quantification of the probabilistic typicality or concentration with respect to the measure \eqref{IDUMO},
\begin{equation}\label{IDTYP}
    \Delta(\rho,A)=\int d(\Pi^j_{B},\rho;A) p_d
\end{equation}
where the probability distribution $p_d$ is obtained from the  $p_j$'s through a Radon-Nikodym derivative. By the Chebyshev inequality, we have that 
\begin{equation}\label{CHEBY}
    P_{\rho}(\abs{\Tr (\Pi^j_{B} A)-\Tr(\rho A)}\geqslant \epsilon)\leqslant\frac{\Delta(\rho,A)^2}{\epsilon^2},\quad\forall\epsilon\in\mathbb{R^+}.
\end{equation}
Therefore, when $ \Delta(\rho,A)$ is very small,  the expectation of the projectively  measured observable would concentrate on the expectation of observable calculated with respect to the state $\rho$. In other words, the indistinguishability of measurement outcomes induces a  description by a mixed state. 

In ETH, one considers the local energy eigenstates. So, let $\sigma^j_{B_1}=\text{Tr}_{\bar{B}_1}\Pi^j_{B}$ be  the reduced projection/state on the subsystem $B_1$. Then we should consider 
\begin{align}\label{BDWSN}
    d(\sigma^j_{B_1},\rho;A^{B_1})=&\abs{\Tr [(\sigma^j_{B_1} -\rho_{B_1}) A^{B_1}]}^2\notag \\
    \leqslant& \norm{\sigma^j_{B_1} -\rho_{B_1}}_1^2 \norm{A^{B_1}}_\infty^2,
\end{align}
where $\norm{\cdot}_{k}$ is the Schatten $k$-norm. 
Next, let $\rho^\text{mc}$ be the density matrix for the microcanonical ensemble.
According to \cref{FLUC,FLUCBVAR,CHEBY}, if 
\begin{equation}\label{14}
\Delta(\rho^\text{mc},A_{B_1})\sim\mathcal{O}(N^{-\alpha}),
\end{equation}
with $0<\alpha<1$,  i.e. the expectations of a local observable $A^{B_1}$ with respect to 
  the results of   local measurements concentrate the expectation of  $A^{B_1}$ with respect to $\rho^\text{mc}$, then we know that each pure state $\sigma^j_{B_i}$ cannot be distinguished from the microcanonical  $\rho^\text{mc}$ in the large $N$ limit. This is the {\it weak ETH with eigenstate typicality} \cite{IKS17}. Furthermore, by using the equivalence of the ensembles, one also has a similar weak ETH on the concentration of  $\sigma^j_{B_i}$ to the canonical-ensemble density matrix $\rho_\text{c}$.
  
  In the proofs of  the weak ETH with eigenstate typicality  for quantum lattice systems \cite{IKS17}, the translation invariance in the following sense is crucial. Let us partition the lattice of system $B$ into $\mathcal{C}=\abs{B}/\abs{B_1}$ blocks with the same size, where $\abs{B}$ means the number lattice points of in $B$. These $\mathcal{C}$ blocks  are   identical copies of $B_1$. Let us also define the translational copies $A^{B_i}$ of $A^{B_1}$ defined on $B_i$ obtained by the translations from block to block. Then the translation invariance means
\begin{equation}\label{TRIV}
    \Tr [\Pi^j_{B} A^{B_i}]=\Tr [\Pi^j_{B} A^{B_1}].
\end{equation}
We can introduce the average observable
\begin{equation}\label{TFLA}
    A^{B}=\frac{1}{\mathcal{C}}\sum_i A^{B_i},
\end{equation}
then the translation invariance \eqref{TRIV} gives 
$\Delta(\rho,A^{B_1})=\Delta(\rho,A^{B})$. Therefore, the weak ETH can be proved  by bounding $\Delta(\rho,A^{B})$. Since the translation invariance of the Hamiltonian does not guarantee the translation invariance of the energy eigenstate, \cref{TRIV} is not unconditionally true for any energy eigenstate and any measurement. If only rely on average observable \cref{TFLA}, then for the translationally invariant state $\rho$, we can also consider
\begin{align}\label{SDTI}
    d(\Pi^j_{B},\rho;A^B)=\abs{\frac{1}{\mathcal{C}}\sum_k\Tr [(\sigma^j_{B_k} -\rho_{B_k}) A^{B_k}]}^2 \notag\\
    =\abs{\Tr [(\frac{1}{\mathcal{C}}\sum_k\sigma^{j,k}_{B_1} -\rho_{B_1}) A^{B_1}]}^2=d(\frac{1}{\mathcal{C}}\sum_k\sigma^{j,k}_{B_1},\rho_{B_1};A^{B_1}),
\end{align}
where $\sigma^{j,k}_{B_1}$ is the translational copies of $\sigma^{j}_{B_k}$. The \cref{SDTI} actually converts the average observable and the average local state into each other. 

\section{ Relating variance to relative entropy}\label{III}
Eqs. \eqref{VAR} and \eqref{IDUMO} depend on the measured observable $A$. In order to quantify the quantum uncertainty in a way that depends only on the quantum measurements but not on the measured observables,  the following entropic uncertainty  used in the entropic uncertainty relation \cite{KLJR14} serves the purpose,
\begin{equation}\label{EUR}
    H_{\Pi}(\rho)=\sum_i p_i S(\rho_i||\rho)
\end{equation}
where $\rho_i =\Pi^i \rho \Pi^i/p_i $ with $p_i=\Tr(\Pi^i \rho)$ and $\{\Pi^i\}$ being the (not necessarily rank-$1$) measurement operators.

In view of the frequent usages of the maps between the total system $B$ and its subsystems $B_i$ in the proofs  in \cite{IKS17}, we consider the  Belavkin-Staszewski (BS) relative entropy \cite{BS82,BCP22}
\begin{align}
    \hat{S}(\sigma||\rho)=&\Tr[\sigma \ln(\mathcal{J}_{\sigma}^{1/2} (\rho^{-1}) )  ]  \label{BS1}\\
    =&\Tr[\sigma \ln(\rho^{-1} \sigma)]    \label{BS3}\\
    =&\Tr[\rho \mathcal{J}_{\rho}^{-1/2} (\sigma) \ln(\mathcal{J}_{\rho}^{-1/2} (\sigma) )  ] \label{BS2}
\end{align}
where $\mathcal{J}^{\alpha}_{\rho}(\cdot):=\rho^\alpha (\cdot)\rho^{\alpha}$ is a rescaling map. Notice that in the above definitions of BS entropy there is the inverse, $\rho^{-1}$, which requires that the density matrix should be strictly positive; this requirement is naturally fulfilled in our considerations, as the density matrices at the position of $\rho$ in the above formulas are the canonical ensemble $\rho^\text{c}$ or the subsystem states.
Now that different $\rho_i$ are orthogonal to each other by definition,  the  entropic uncertainty can be generalized by using \eqref{BS1} as
\begin{equation}
    \sum_i p_i \hat{S}(\rho_i||\rho)=\Tr [\rho \ln(\sum_i \mathcal{J}_{\rho_i}^{1/2} (\rho^{-1}))].
\end{equation}
When the Hilbert-Schmidt norm $\norm{X-I}_2\leqslant 1$, the power series of the matrix logarithm 
\begin{equation}
    \ln(X)=(X-I)-\frac{1}{2}(X-I)^2+\dots
\end{equation}
converges.
Using it,  we can obtain the following first-order relations
\begin{align}\label{RETVAR}
    &\sum_i \Tr \rho [ \mathcal{J}_{\rho_i}^{1/2} (\rho^{-1})-1] = \sum_i p_i\Bigl( \Tr \rho_i  [\mathcal{J}_{\rho_i}^{1/2} (\rho^{-1})]-1\Bigr)\notag \\
    = &\sum_i p_i\Bigl(\Tr [\rho  (\mathcal{J}_{\rho}^{-1/2} (\rho_i))^2]-1\Bigr)=\sum_i p_i V(\rho,O_i)
\end{align}
where in the first line we have used  $\Tr\rho_i=1$,
 the second line follows from \eqref{BS2}, and 
 \begin{equation}\label{OOO}
 O_i:=\mathcal{J}_{\rho}^{-1/2} (\rho_i)
 \end{equation}
 with $\braket{O_i}=1$. This $O_i$ plays the role of observable in quantum variance, and it is defined by $\rho_i$ in a one-to-one manner. Although an observable $O$ can be mathematically related to a particular density matrix $\rho$, the physical meaning of such an $O$ is possibly unclear. Therefore, we do not interpret this $O_i$ in \eqref{OOO} and merely take it as an intermediate technical step. Formally,
Eq. \eqref{RETVAR} establishes a link between the variance of $O_i$ in the states $\rho$ and the entropic uncertainty in the first-order sense. 
Since the quantum relative entropy encodes the closeness between two density matrices, the $V(\rho,O_i)$ is again a quantity measuring the (in)distinguishability between state $\rho_i$ and $\rho$.

The relation \eqref{RETVAR} is suitable   for studying localized states on subsystems.
Let $\rho_B=\sum_j p_j \Pi^j_{B}$ as before. For the pure state $\Pi^j_{B}$, its reduced density matrix on a subsystem, say $B_1$, $\sigma^j_{B_1}=\text{Tr}_{\bar{B}_1}\Pi^j_{B}$ is no longer pure in general, so that it can be arbitrary subsystem states of $B_1$. The reduced density matrix of $\rho_B$ on $B_1$ is $\rho_{B_1}=\sum_i p_i \sigma^i_{B_1}$.
In this setting, we can  consider the {\it formal} observable
\begin{equation}
    O^{B_1}_i=\mathcal{J}_{\rho_{B_1}}^{-1/2} (\sigma^i_{B_1}).
\end{equation}
Again, we have  $\braket{O^{B_1}_i}=1$. Similar to \eqref{RETVAR}, we also have
\begin{align}\label{VARTRE}
    V(\rho_B,O^{B_1}_i)=&\Tr \rho_B  [(\mathcal{J}_{\rho_{B_1}}^{-1/2} (\sigma^i_{B_1}))^2-1]=\notag \\
    =&\Tr [\sigma^i_{B_1} (\mathcal{J}_{\sigma^i_{B_1}}^{1/2} (\rho_{B_1}^{-1})-1)]
\end{align}
as the first order expansions of 
$\hat{S}(\sigma^i_{B_1} ||\rho_{B_1})$. In \eqref{VARTRE} we have used the property that $\text{Tr}\rho_{B}=1$ as a normalized density matrix.
Eq. \eqref{VARTRE} relates the  indistinguishability of localized states and the measurement uncertainty (of $O^{B_1}_i$) in $\rho$, in an average sense.

Recall that the BS relative entropy and the quantum relative entropy satisfy $  \hat{S}(\sigma||\rho)\geqslant S(\sigma||\rho) $ \cite{BCP22}, thereby
\begin{equation}\label{28}
    \hat{S}(\sigma^i_{B_1} ||\rho_{B_1})\geqslant\frac{1}{2}\norm{\sigma^i_{B_1}-\rho_{B_1}}_1^2
\end{equation}
where the Schatten-$1$ norm $\norm{\cdot}_1$ is just the trace distance introduced above. On the other hand, the variance $ V(\rho_B,O^{B_1}_i)$ before the series expansion is by definition a Schatten-$2$ norm,
\begin{align}\label{VTS2N}
    V(\rho_B,O^{B_1}_i)&=\Tr[(\sigma^i_{B_1}-\rho_{B_1})\rho_{B_1}^{-1}(\sigma^i_{B_1}-\rho_{B_1})]=\notag\\
     &=\norm{(\sigma^i_{B_1}-\rho_{B_1})\rho^{-1/2}_{B_1}}_2^2.
\end{align}
By  H\"{o}lder's inequality, we have 
\begin{align}
    \norm{(\sigma^i_{B_1}-\rho_{B_1})}^2_1=& \norm{(\sigma^i_{B_1}-\rho_{B_1})\rho^{-1/2}_{B_1}\rho^{1/2}_{B_1}}^2_1\notag \\
    \leqslant& \norm{(\sigma^i_{B_1}-\rho_{B_1})\rho^{-1/2}_{B_1}}^2_2\norm{\rho^{1/2}_{B_1}}^2_2\notag\\
    =&\norm{(\sigma^i_{B_1}-\rho_{B_1})\rho^{-1/2}_{B_1}}^2_2=V(\rho_B,O^{B_1}_i).\label{30}
\end{align}

Similarly, we can consider the ``off-diagonal'' observable
\begin{equation}\label{OFFDGO}
    O^{B_1}_{ij}=\mathcal{J}_{\rho_{B_1}}^{-1/2} (\sigma^{ij}_{B_1}),\quad i\neq j,
\end{equation}
with $\sigma^{ij}_{B_1}=\text{Tr}_{\bar{B}_1}\Pi^{ij}_{B}$ an ``off-diagonal'' reduced density matrix. Now we have $\braket{O^{B_1}_{ij}}=0$. Again,  we have
\begin{align}
    \norm{\sigma^{ij}_{B_1}}^2_1=& \norm{\sigma^{ij}_{B_1}\rho^{-1/2}_{B_1}\rho^{1/2}_{B_1}}^2_1\notag \\
    \leqslant& \norm{\sigma^{ij}_{B_1}\rho^{-1/2}_{B_1}}^2_2\norm{\rho^{1/2}_{B_1}}^2_2\notag\\
    =&\norm{\sigma^{ij}_{B_1}\rho^{-1/2}_{B_1}}^2_2=V(\rho_B,O^{B_1}_{ij}).\label{323232}
\end{align}
where the second line follows from H\"{o}lder's inequality and the third line holds by definition.

Similar to the definition
 \eqref{BS3}, we can also rewrite the variance \eqref{VARTRE}  as 
\begin{equation}\label{VARTRE3}
    \Tr [  (\mathcal{J}_{\rho_{B}}^{1/2}\circ\mathcal{J}_{\rho_{B_1}}^{-1/2} (\sigma^i_{B_1}))(\rho_{B}^{-1}\mathcal{J}_{\rho_{B}}^{1/2}\circ\mathcal{J}_{\rho_{B_1}}^{-1/2} (\sigma^i_{B_1})-1)],
\end{equation}
which is the first-order expansion of 
$\hat{S}( \mathcal{J}_{\rho_{B}}^{1/2}\circ\mathcal{J}_{\rho_{B_1}}^{-1/2} (\sigma^i_{B_1}) ||\rho_{B})$. In this form \eqref{VARTRE3}, we find that the map
\begin{equation}
   \mathcal{R}_{\rho}^{B_k\to B}=\mathcal{J}_{\rho_{B}}^{1/2}\circ\mathcal{J}_{\rho_{B_k}}^{-1/2} 
\end{equation}
is just the Petz recovery map of the completely positive trace-preserving (CPTP) map $\mathcal{N}_{B\to B_k}=\Tr_{\bar{B}_k}$ with respect to the reference state $\rho_B$, cf. \cite{KK19}. In this way, we can rewrite, by using \cref{BS3,BS2},
\begin{align}\label{LOCID}
    \hat{S}(\sigma^i_{B_1} ||\rho_{B_1})=&\Tr[\rho_{B_1} \mathcal{J}_{\rho_{B_1}}^{-1/2} (\sigma^i_{B_1}) \ln(\mathcal{J}_{\rho_{B_1}}^{-1/2} (\sigma^i_{B_1}) )  ] \notag\\
    =&\Tr[\mathcal{R}_{\rho}^{B_1\to B} (\sigma^i_{B_1})  \ln(\rho_{B}^{-1}\mathcal{R}_{\rho}^{B_1\to B} (\sigma^i_{B_1}) )] \notag\\
     =&\hat{S}(\mathcal{R}_{\rho}^{B_1\to B} (\sigma^i_{B_1}) ||\rho_{B}).
\end{align}
The final expression pulls the subsystem BS entropy to the global one which would be easier to make  bounds.

A thing we should keep in mind is that the relations derived in this section are mainly mathematical relations with their physical meanings uninterpreted. The punchline is that 
 we can approach the subsystem ETH \eqref{3} and \eqref{4}
 by bounding either $V(\rho_B,O^{B_1}_i)$, or $\hat{S}(\mathcal{R}_{\rho}^{B_1\to B} (\sigma^i_{B_1}) ||\rho_{B})$, and $V(\rho_B,O^{B_1}_{ij})$ based on \eqref{28}, \eqref{30}, \eqref{323232}, and \eqref{LOCID}. 
\section{Subsystem ETH for translation invariant systems}\label{IV}
Now we can turn to  the proof of the subsystem ETH. The strategy is to derive general bounds on the trace distance and show that they are small in the large $N$ limit. 

We consider  the macroscopic observable that is composed solely of local operators as in \cite{KS20}, or the translation invariant quantum lattice systems as in the last paragraph of section \ref{IIB} of \cite{IKS17}.
As in \eqref{TFLA}, we define the average {\it formal} observable
\begin{equation}\label{AFO}
    O^{B}_i=\frac{1}{\mathcal{C}}\sum_{k}O^{B_k}_i=\frac{1}{\mathcal{C}}\sum_{k}\mathcal{J}_{\rho_{B_k}}^{-1/2} (\sigma^{i,1}_{B_k}),
\end{equation}
where $\sigma^{i,1}_{B_k}$ is the translational copies of $\sigma^{i}_{B_1}$. It can also be obtained by translating state $\Pi^i_B$ and then taking the partial trace.
Here, we assume an equipartition of the lattice into subsystems with the same size, so that 
\begin{equation}\mathcal{C}=N/N_A\end{equation}
if the number of sites in $B_1$ is $N_A$.
We still have $\braket{O^{B}_i}=1$. The quantum variance
\begin{align}\label{TIVAR}
    &V(\rho,O^{B}_i)= \notag\\
    =&\Tr \Bigl[  (\frac{1}{\mathcal{C}}\sum_{k}\mathcal{R}_{\rho}^{B_k\to B} (\sigma^{i,1}_{B_k}))(\rho_B^{-1}\frac{1}{\mathcal{C}}\sum_{l}\mathcal{R}_{\rho}^{B_l\to B} (\sigma^{i,1}_{B_l})-1)\Bigr],
\end{align}
as given by \cref{VARTRE3}, is the first order expansion of the BS relative entropy
\begin{equation}\label{LOCIVID}
    \hat{S}\Bigl(\frac{1}{\mathcal{C}}\sum_{k}\mathcal{R}_{\rho}^{B_k\to B} (\sigma^{i,1}_{B_k}) ||\rho_{B}\Bigr).
\end{equation}
Since  the Petz recovery map $\mathcal{R}_{\rho}^{B_k\to B}$ is also CPTP, we see that 
$  \frac{1}{\mathcal{C}}\sum_{k}\mathcal{R}_{\rho}^{B_k\to B}(\sigma^{i,1}_{B_k})$ is also a legitimate density matrix. For example, consider that there is no correlation between the blocks  $B_1,\dots,B_{\mathcal{C}}$, i.e. $\rho_B=\rho_{B_1}\otimes \dots \otimes \rho_{B_\mathcal{C}}$, then
\[
    \mathcal{R}_{\rho}^{B_k\to B} (\sigma^{i,1}_{B_k})=\rho_{B_1}\otimes \dots \otimes \sigma^{i,1}_{B_k} \otimes \dots \otimes \rho_{B_\mathcal{C}}.
\]
By the joint convexity of relative entropy, it is easy to show that 
\begin{align}
    \hat{S}\Bigl(\frac{1}{\mathcal{C}}\sum_{k}\mathcal{R}_{\rho}^{B_k\to B} (\sigma^{i,1}_{B_k}) ||\rho_{B}\Bigr)&\leqslant \frac{1}{\mathcal{C}}\sum_{k} \hat{S}(\mathcal{R}_{\rho}^{B_k\to B} (\sigma^{i,1}_{B_k}) ||\rho_{B})\notag \\
    =\frac{1}{\mathcal{C}}\sum_{k} \hat{S}(\sigma^{i,1}_{B_k} ||\rho_{B_k})&=\hat{S}(\sigma_{B_1} ||\rho_{B_1}),\label{35}
\end{align}
the last expression of which  is just the local (in)distinguishability. In \eqref{35} we supposed that the state $\rho_B$ is translation invariant; this requirement is naturally fulfilled by the canonical ensemble.  As we can see from \eqref{35}, if  the $\hat{S}(\sigma^{i,1}_{B_k} ||\rho_{B_k})$ are small for all blocks $B_i$, then $ \hat{S}(\frac{1}{\mathcal{C}}\sum_{k}\mathcal{R}_{\rho}^{B_k\to B} (\sigma^{i,1}_{B_k}) ||\rho_{B})$ must be small, but the converse is not true. 

To prove the subsystem ETH, we need to show that 
\begin{align}
\hat{S}\Bigl(\frac{1}{\mathcal{C}}\sum_{k}\mathcal{R}_{\rho}^{B_k\to B} (\sigma^{i,1}_{B_k}) ||\rho^\text{c}_{B}\Bigr)&\sim\mathcal{O}({N_A/N}),\label{36}\\
 \text{or}\quad V(\rho^\text{c},O^{B}_i)&\sim\mathcal{O}({N_A/N}),\label{37}
\end{align}

Firstly, the quantum variance can be rewritten as
\begin{align}
    V(\rho,O^{B}_i)&=\frac{1}{\mathcal{C}^2}\sum_{k} V(\rho,O^{B_k}_i)+\notag \\
    &+\frac{1}{\mathcal{C}^2}\sum_{k\neq l}\Tr [O^{B_k}_i\otimes O^{B_l}_i(\rho_{B_kB_l}-\rho_{B_k}\otimes\rho_{B_l}) ].\label{26}
\end{align}
The first term in \eqref{26} is the local variance, in which the terms
\begin{equation}
    V(\rho,O^{B_k}_i)=\norm{(\sigma^{i,1}_{B_k}-\rho_{B_k})\rho^{-1/2}_{B_k}}_2^2=V(\rho,O^{B_1}_i)
\end{equation}
will not grow with $\mathcal{C}$. So we have
\begin{equation}\label{39}
\frac{1}{\mathcal{C}^2}\sum_{k} V(\rho,O^{B_k}_i)= V(\rho,O^{B_1}_i)\times\mathcal{C}^{-1}.\end{equation}
The second term in \eqref{26} depends on the correlations between $B_k$ and $B_l$.
 Suppose that the correlations of the canonical thermal state decay algebraically, i.e. 
 \begin{equation}\label{CCDC}
    \norm{\rho^\text{c}_{B_kB_l}-\rho^\text{c}_{B_k}\otimes\rho^\text{c}_{B_l}}\leqslant d(B_k,B_l)^{-\gamma},\quad\gamma\geqslant D_L
\end{equation}
 where $D_L$ is spatial dimension of the lattice and $d(A,B)$ is the shortest lattice path length between two regions $A$ and $B$. The $\gamma$ characterizes the decay of the correlations, which is related to the specific model. 
 Then the term in the second term of \eqref{26} is less than or equal to
 \begin{equation}\label{SR}
 \frac{O^2_{\max}}{\mathcal{C}^2}\sum_{k} \sum_{d=1}^{\infty} n_d d^{-\gamma}=d^{-\gamma}_{\text{eff}} \times \frac{O^2_{\max}}{\mathcal{C}},
 \end{equation}
 where $O_{\max}=\norm{O^{B_1}_i}_\infty$ and $n_d$ is the number of blocks that are of distance $d$ from $B_k$. For lattices with spatial dimension $D_L$, we have in general $n_d\propto d^{D_L-1}$.  The
 $d_{\text{eff}}$ is the effective distance given by $ \sum_{d=1}^{\infty} n_d d^{-\gamma}$, while the $\sum_k$ in \eqref{SR} gives $\mathcal{C}$.
Combining the above bounds, we see that \eqref{37} holds.
Due to the translation invariance of $\rho^\text{c}$, the variance for different blocks should give the same result. 
Therefore, for many $O^{B}_i$ or equivalent $\sigma^i$, there should be
\begin{equation}\label{TIVVR}
    V(\rho,O^{B}_i)\sim  V(\rho,O^{B_k}_i)=V(\rho,O^{B_1}_i)
\end{equation}
This is an analog of the relation $\Delta(\rho,A^{B_1})=\Delta(\rho,A^{B})$ below \eqref{TFLA}, since the $O_i^{B_k}$ is also the translational copies of $O_i^{B_1}$ according to \cref{AFO}.
With (\ref{TIVVR}), we can rewrite \cref{26} as
\begin{equation}\label{BFV}
   V(\rho,O^{B}_i)\sim\frac{1}{\mathcal{C}(\mathcal{C}-1)}\sum_{k\neq l}\Tr [O^{B_k}_i\otimes O^{B_l}_i(\rho_{B_kB_l}-\rho_{B_k}\otimes\rho_{B_l}) ].
\end{equation}
It can provide a slightly tighter bound.


Secondly, we study the bounds on the BS relative entropy \eqref{LOCIVID}. To this end, define 
the $m$-th moment of the (expanded logarithm) operator $O^{B}_i-1$,
\begin{equation}
    M^{(m)}=\Tr[\rho^\text{c}_B (O^{B}_i-I)^{m}]
\end{equation}
which is the higher-moment generalization of \eqref{VARTRE}. Then, by the power series of the matrix logarithm, we have 
\begin{align}
    &\hat{S}\Bigl(\frac{1}{\mathcal{C}}\sum_{k}\mathcal{R}_{\rho}^{B_k\to B} (\sigma^{i,1}_{B_k}) ||\rho^\text{c}_{B}\Bigr)=\notag \\
   =&\frac{1}{\mathcal{C}}\sum_{k} \Tr[\rho^\text{c}_B O^{B_k} \ln(\frac{1}{\mathcal{C}}\sum_{l}O^{B_l})  ] = \notag\\
    =&V(\rho^\text{c},O^{B}_i)+\dots+\frac{(-1)^{n}}{n-1}(M^{(n)}+ M^{(n-1)})+\dots\notag\\
    =&\frac{1}{2}V(\rho^\text{c},O^{B}_i)+\sum_{n=3}^{\infty}\frac{(-1)^{n}}{(n-1)n}M^{(n)}.\label{43}
\end{align}
The first term $ V(\rho^\text{c},O^{B}_i)$ has been bounded as in \eqref{SR}. The other terms in \eqref{43} depend on the multipartite correlations, and the higher moments $M^{(m)}$ in them can be bounded in the same way as in  \cite{KS20},
\begin{equation}
    M^{(m)}\leqslant \frac{1}{\mathcal{C}^m}\mathcal{O}(\mathcal{C}^{m/2})\sim\mathcal{O}(\mathcal{C}^{-m/2}),
\end{equation}
where we omit those parts that do not increase with $\mathcal{C}$.
This $\mathcal{O}(\mathcal{C}^{-m/2})$ behavior decays faster than $\mathcal{O}(\mathcal{C}^{-1})$, so the BS entropy should be mainly bound by the behavior of the first variance term.
Thus, we obtain the overall bounds \eqref{36} on the BS relative entropy.

Thirdly, if we replace the canonical thermal state $\rho^\text{c}$ by another local state $\sigma_{B_l}$ in the above formulas, we will find that the scaling analysis still holds. In other words, for two local states, we have 
\begin{equation}\label{4949}
\norm{\sigma^i_{B_k}-\sigma^i_{B_l}}\sim \mathcal{O}(\mathcal{C}^{-1/2}).
\end{equation}
This means the concentration of states of different subsystems to certain common equilibrium state. 
However, this \eqref{4949} is not the ``off-diagonal'' subsystem ETH \eqref{4}. In fact, \eqref{4} holds in the following sense: From the inequality \eqref{323232} and the fact that  the bound on variance in the first step of proof does not depend on the specific forms of measurements, one obtains
\begin{equation}
\norm{\sigma^{ij}}_1\sim\mathcal{O}(\mathcal{C}^{-1/2}).
\end{equation}

We have therefore successfully proved the subsystem ETH by showing the bounds or decaying behaviors \eqref{36} and \eqref{37}.  We remark that this decay behavior $\mathcal{O}(N_A^{1/2}/N^{1/2})$ is qualitatively consistent with the observations made in \cite{DLL18} that the subsystem must be small compared to the total system size.  This is simply because for larger $N_A$ the faster the decaying speed, and hence the remaining bound should be the smaller $N_A$.

In the previous proof, we mainly considered the case where $\rho_B$ is a Gibbs state. But in fact, our proof mainly uses the strict positivity of $\rho_B$ and the correlation decay \cref{CCDC}. Therefore, as long as these two properties are satisfied, other states can also be used. Such as microcanonical ensembles or certain evolutionary steady states. Of course, when other states are selected, the bounds of $O_{\max}$ will also be affected, thus affecting the tightness of the bound.


Compared to the proofs given in (the appendix) of \cite{IKS17}, we have changed the (in)distinguishability measure \eqref{IDUMO} to the variance or BS relative entropy.
We can apply such a replacement back to the proofs of the weak ETH with eigenstate typicality as in \cite{IKS17} to see what happens. Similar to \cref{BDWSN,30}, the (in)distinguishability measure in \cref{SDTI} satisfies
\begin{equation}
    \abs{\Tr [(\Pi^i -\rho_{B}) A^{B}]}^2\leqslant V(\rho,\frac{1}{\mathcal{C}}\sum_{k}\mathcal{J}_{\rho_{B_1}}^{-1/2} (\sigma^{i,k}_{B_1}))\norm{A^{B_1}}_\infty^2.
\end{equation}
By replacing $d(\Pi^j_{B},\rho;A)$ with the variance, we see that the probabilistic typicality  \eqref{IDTYP} becomes
\begin{equation}\label{CAQV}
   \braket{V_\text{dg}}:= \sum_{i} p_iV(\rho,\frac{1}{\mathcal{C}}\sum_{k}\mathcal{J}_{\rho_{B_1}}^{-1/2} (\sigma^{i,k}_{B_1})).
\end{equation}
Similarly, we can consider the ``off-diagonal'' probabilistic typicality
\begin{equation}\label{CAQVOFF}
    \braket{V_\text{off}}:= \sum_{i\neq j} p_iV(\rho,\frac{1}{\mathcal{C}}\sum_{k}\mathcal{J}_{\rho_{B_1}}^{-1/2} (\sigma^{ij,k}_{B_1})).
 \end{equation}
 Similar to \cref{BDWSN,30}, the off-diagonal measure also satisfies
 \begin{equation}
    \abs{\Tr [\Pi^{ij}_{B}  A^{B}]}^2\leqslant V(\rho,\frac{1}{\mathcal{C}}\sum_{k}\mathcal{J}_{\rho_{B_1}}^{-1/2} (\sigma^{ij,k}_{B_1}))\norm{A^{B_1}}_\infty^2, \quad i\neq j.
\end{equation}
Let $\rho_{B_1}=\sum_\alpha p'_\alpha \Pi^\alpha_{B_1}$ be the state of the subsystem $B_1$ expanded in the orthonormal basis $\{\Pi^\alpha_{B_1}\}$ of rank-$1$ projectors. With these projectors and \cref{VTS2N,323232}, one can rewrite \cref{CAQV,CAQVOFF} as 
\begin{align}\label{DGODG}
    \braket{V_\text{dg}}+  \braket{V_\text{off}}= \sum_{i,\alpha,\beta} p_i {p'_\alpha}^{-1}\abs{\Tr[(\frac{1}{\mathcal{C}}\sum_{k}\sigma^{i,k}_{B_1}-\rho_{B_1})\Pi^{\alpha \beta}_{B_1}]}^2\notag\\
    +\sum_{i\neq j,\alpha,\beta } p_i {p'_\alpha}^{-1}\abs{\Tr[(\frac{1}{\mathcal{C}}\sum_{k}\sigma^{ij,k}_{B_1})\Pi^{\alpha \beta}_{B_1}]}^2.
\end{align}
Notice that the transformation (\ref{SDTI}) also applies to off-diagonal terms
\begin{equation}\label{SDTIOFF}
   \abs{\Tr [(\frac{1}{\mathcal{C}}\sum_k\sigma^{ij,k}_{B_1})  A^{B_1}]}^2=\abs{\Tr [\Pi^{ij} (\frac{1}{\mathcal{C}}\sum_k A^{B_k})]}^2,  \quad i\neq j
\end{equation}
Using \cref{SDTI,SDTIOFF}, we can convert the average local state back to the average observable. Then according to the form of variance in formula (\ref{FLUCBVAR}), we have
\begin{equation}\label{AVVCAN}
    \braket{V_\text{dg}}+  \braket{V_\text{off}}= \sum_{\beta, \alpha} p'_\beta V(\rho,\frac{1}{\mathcal{C}}\sum_{k}\mathcal{J}_{\rho_{B_k}}^{-1/2} (\Pi^{\alpha \beta}_{B_k})),
\end{equation}
where $\Pi^{\alpha \beta}_{B_k}$ is the translational copies of $\Pi^{\alpha \beta}_{B_1}$. In \eqref{AVVCAN} we have used the property that
\begin{equation}
    \mathcal{J}_{\rho_{B_k}}^{-1/2} (\Pi^{\alpha \beta}_{B_k})=(p'_\alpha p'_\beta)^{-1/2}\Pi^{\alpha \beta}_{B_k}.
\end{equation}
Since we assume that state $\rho$ is translation invariant, therefore $\Pi^{\alpha}_{B_k}$ is still the diagonal basis of $\rho_{B_k}$. The right-hand side of inequality (\ref{AVVCAN}) can be bounded like inequality (\ref{26}). It should be pointed out that due to the orthogonal relationship between operators
\begin{align}
    \mathcal{J}_{\rho_{B_k}}^{-1/2} (\Pi^{\alpha}_{B_k})[\mathcal{J}_{\rho_{B_k}}^{-1/2} (\Pi^{\alpha \beta}_{B_k})]^\dagger=0 ,\quad \beta \neq \alpha\notag\\
    \mathcal{J}_{\rho_{B_k}}^{-1/2} (\Pi^{\alpha \gamma}_{B_k})[\mathcal{J}_{\rho_{B_k}}^{-1/2} (\Pi^{\alpha \beta}_{B_k})]^\dagger=0 ,\quad \beta \neq \gamma,
\end{align}
the corresponding local variance term satisfies
\begin{align}
    p'_\alpha V(\rho,\mathcal{J}_{\rho_{B_k}}^{-1/2} (\Pi^{\alpha}_{B_k}))+ \sum_{\beta, \beta \neq \alpha}p'_\beta V(\rho,\mathcal{J}_{\rho_{B_k}}^{-1/2} (\Pi^{\alpha \beta}_{B_k}))\notag\\
    = p'_\alpha V(\rho, {p'_\alpha}^{-1}(\Pi^{\alpha}_{B_k}+\sum_{\beta, \beta\neq \alpha}\Pi^{\alpha \beta}_{B_k})).
\end{align}
The other terms are very small as long as the correlations decay fast enough. Combining \cref{CAQV,CAQVOFF,AVVCAN}, we have the Chebyshev-type inequality, 
\begin{align}\label{LOCIDMCTYP}
    &P_{\rho}(\sum_{j} V(\rho,\frac{1}{\mathcal{C}}\sum_{k}\mathcal{J}_{\rho_{B_1}}^{-1/2} (\sigma^{ij,k}_{B_1}))\geqslant \epsilon^2)\notag\\
    &\leqslant \frac{1}{\epsilon^2}\left[\sum_{ \beta,\alpha} p'_\beta V(\rho,\frac{1}{\mathcal{C}}\sum_{k}\mathcal{J}_{\rho_{B_k}}^{-1/2} (\Pi^{\alpha \beta}_{B_k}))\right]
\end{align}
for $\epsilon>0$.
When  the right-hand side of \cref{LOCIDMCTYP} is very small, we can conclude that  the measurement results concentrate on the results predicted by $\rho$. It is similar to the weak ETH with eigenstate typicality, but it includes both diagonal and off-diagonal ETH and does not depend on specific measurements.

The local observable in $V_\text{dg}$ and $V_\text{off}$ only measure the state of $B_1$, but it is determined by the average state of each block. On the contrary, the observable (\ref{AFO}) will measure the state of each block, but is only determined by the state of $B_1$.  They look very different, but they are deeply connected, as we will show below. 
In \cref{DGODG}, we use the variation form (\ref{VTS2N}) and the spectral decomposition of $\rho_{B_1}$. If we use the variation form (\ref{VARTRE3}) and the spectral decomposition of $\rho_{B}$ instead, we get
\begin{align}
    \braket{V_\text{off}}+\braket{V_\text{dg}}= \sum_{i,j} p_i V(\rho,\frac{1}{\mathcal{C}}\sum_{k}\mathcal{J}_{\rho_{B_1}}^{-1/2} (\sigma^{ij,k}_{B_1}))\notag\\
    =\sum_{i\neq j,\alpha,\beta }p_i {p_\alpha}\abs{\Tr[\mathcal{J}_{\rho_{B_1}}^{-1/2}(\frac{1}{\mathcal{C}}\sum_{k}\sigma^{ij,k}_{B_1})\sigma^{\alpha \beta}_{B_1}]}^2 \notag\\
    +\sum_{i,\alpha,\beta} p_i {p_\alpha}\abs{\Tr[\mathcal{J}_{\rho_{B_1}}^{-1/2}(\frac{1}{\mathcal{C}}\sum_{k}\sigma^{i,k}_{B_1}-\rho_{B_1})\sigma^{\alpha \beta}_{B_1}]}^2\notag\\
    =  \sum_{i, j,\alpha\neq\beta } p_i {p_\alpha}\abs{\Tr[\frac{1}{\mathcal{C}}\sum_{k}\sigma^{ij}_{B_k}\mathcal{J}_{\rho_{B_k}}^{-1/2}(\sigma^{\alpha \beta,1}_{B_k})]}^2 \notag\\
    +\sum_{i,j,\alpha} p_i {p_\alpha}\abs{\Tr[\frac{1}{\mathcal{C}}\sum_{k}\sigma^{ij}_{B_k}\mathcal{J}_{\rho_{B_k}}^{-1/2}(\sigma^{\alpha,1}_{B_k}-\rho_{B_k})]}^2\notag\\
    =\sum_{\alpha, \beta} p_\alpha V(\rho,\frac{1}{\mathcal{C}}\sum_{k}\mathcal{J}_{\rho_{B_k}}^{-1/2} (\sigma^{\alpha \beta,1}_{B_k})).
\end{align}
This equation establishes the connection between  $V_\text{dg}$, $V_\text{off}$ and $V(\rho,O^{B}_i)$, $V(\rho,O^{B}_{i\neq j})$.

Now we briefly discuss the equivalence between the microcanonical and canonical ensembles. To this end, we consider a microcanonical energy shell $(E-\delta,E]$ with width $\delta$ with the index set
\begin{equation}
\mathcal{M}_{E,\delta}=\{i| E_i\in(E-\delta,E]\}.
\end{equation}
The (in)distinguishability of the microcanonical and canonical ensembles can be bounded with
\begin{align}\label{47}
    \norm{\rho^\text{mc}_{B_1} -\rho^\text{c}_{B_1}}_1\leqslant  \sum_{i\in \mathcal{M}_{E,\delta}} \frac{1}{D}\norm{\frac{1}{\mathcal{C}}\sum_k\sigma^{i,k}_{B_1}-\rho^\text{c}_{B_1}}_1 \notag\\
    \leq  \sum_{i\in \mathcal{M}_{E,\delta}}\frac{1}{D}\left[V(\rho^\text{c},\frac{1}{\mathcal{C}}\sum_{k}\mathcal{J}_{\rho^\text{c}_{B_1}}^{-1/2} (\sigma^{i,k}_{B_1}))\right]^{1/2}
\end{align}
where we have used the joint convexity of Schatten norm and \cref{30}. In the large $N$ limit, we have from \eqref{LOCIDMCTYP} that  the right-hand side of \eqref{47} is very small, so we can conclude the equivalence between the microcanonical and canonical ensembles  in this case. 


\section{The bound from the clustering of correlations}\label{5V}

It seems that the Hamiltonian of the system does not make an appearance in the above proof, but in fact, the Hamiltonian is important in the condition \eqref{CCDC} of  correlations leading to \eqref{SR}.  We see that as long as the correlations decay fast enough, i.e. $\gamma\geqslant D_L$, the scaling \eqref{SR} and hence the above proof of the subsystem ETH holds. For models with  exponentially decaying correlations, the above conditions can be easily satisfied.  

We remark that the behavior of \eqref{SR} holds not only for short-range interactions, but also for some types of long-range interactions. To see this, let us recall that the mutual information of the Gibbs state has some general bounds; in particular, for long-range interactions of the form $1/d^{\eta+D_L},\eta>0$, we have for high temperatures the following bound on mutual information between two regions $A$ and $C$,
\begin{equation}
    I(A:C)\leqslant \beta \min(N_{A},N_{C})\frac{C_\beta}{d(A,C)^{\eta}}
\end{equation}
where $C_\beta$ is a function of the inverse temperature $\beta$ independent on the system size which can be found in \cite{KKB20}. The mutual information can be related to the relative entropy through $I(A:C)=S(\rho_{AC}||\rho_A\otimes\rho_{C})$, whence
\begin{equation}
     \norm{\rho^\text{c}_{B_kB_l}-\rho^\text{c}_{B_k}\otimes\rho^\text{c}_{B_l}}\leqslant \sqrt{2I(B_k:B_l)}\leqslant \frac{N^{1/2}_A\sqrt{2\beta C_\beta}}{d(B_k,B_l)^{\eta/2}}
\end{equation}
where we have assumed $N_A<N_C$ and used \eqref{28}.
Since $n_d\propto d^{D_L-1}$, we obtain
\begin{equation}\label{CORDIVB}
   \lim_{\mathcal{C}\to \infty} \sum_{d=1}^{\mathcal{C}^{{1}/{D_L}}} d^{D_L-1} d^{-\eta/2}  \times(\sum_{k}\frac{1}{\mathcal{C}^2}) =O( \mathcal{C}^{-\eta/(2D_L)}).
\end{equation}
When $\eta>0$, it is possible that  the estimate \eqref{SR} still holds. We see that, for one-dimensional systems ($D_L=1$), we require $\eta=2$ to conform to the estimate \eqref{SR}. Compared to 
the numerical results reported in \cite{SHU22}, this value is within the range of validity of strong ETH, i.e. $\eta+D_L\geqslant0.6$, although with a slower speed of convergence.



\section{Conclusion and discussion}\label{CD}
We have studied the subsystem ETH for translation invariant  quantum  systems. We develop upon the setting for translation invariant systems given in \cite{IKS17} by relating the quantum variance to the BS relative entropy. Surprisingly, with this technical input, we are able to prove the subsystem ETH for translation invariant systems using the similar scaling analysis as in \cite{IKS17}. The proof given above is elementary, without referring to the advanced techniques from random matrix theory. Since the subsystem is stronger than the local ETH, our results corroborate the previous results for local ETH for translation invariant systems \cite{SE,Mori,IKS17}.


We have remarked that our results apply to some long-range interacting systems. Compared with  the recent numerical test for one-dimensional translation invariant systems \cite{SHU22}, the constraint on the interaction parameter here is less stringent, but can be applied to other dimensions.
However, adding an external driving field will make the system going nonequilibrium \cite{RVS23}, even when the system is translation invariant. 
Another point is that our results only restrict the decaying of error terms to be algebraically. The exponential decays of errors is a quite  strong results, which might not be universal in view of the examples  from large-$c$ CFTs with $\mathcal{O}({c}^0)$ decay \cite{HLZ17a,HLZ17b,GLZ19}.

 In the  analysis of \eqref{43} the higher moments are relevant. The higher-moment versions of ETH can be related to many interesting structures, such as the out-of-time-ordered correlation functions indicating  quantum chaos \cite{FK19}. This could be a possible approach to relating the chaotic conjecture and the present analysis without referring to random matrices. 
Moreover,  it is also interesting to study the eigenstate fluctuation theorems \cite{IKS17,IKS22} at the subsystem level, which might be a suitable situation for thermalized open quantum systems.
These aspects are left to future investigations.
\begin{acknowledgments}
We would like to thank Anatoly Dymarsky, Qiang Miao, and  J\"{u}rgen Schnack for their helpful comments.
    This work is supported by the National Natural Science Foundation of
China under Grant No. 12305035. 
\end{acknowledgments}

\end{CJK*}

\end{document}